\documentclass[preprint]{aastex}
\usepackage{multirow}
\shorttitle{Secular Gravitational Instability}
\shortauthors{Michikoshi et al.}

\begin{document}
\title{Secular Gravitational Instability of a Dust Layer in Shear Turbulence}

\author{Shugo Michikoshi\altaffilmark{1}, 
Eiichiro Kokubo\altaffilmark{1,2}, and
Shu-ichiro Inutsuka\altaffilmark{3} 
}
\altaffiltext{1}{
Center for Computational Astrophysics, National Astronomical Observatory of Japan, Osawa, Mitaka, Tokyo 181-8588, Japan
}
\altaffiltext{2}{
Division of Theoretical Astronomy, National Astronomical Observatory of Japan, Osawa, Mitaka, Tokyo 181-8588, Japan
}
\altaffiltext{3}{
Department of Physics, Graduate School of Science, Nagoya University, Furo-cho, Chikusa-ku, Nagoya, Aichi 464-8602, Japan
}
\email{michikoshi@cfca.jp, kokubo@th.nao.ac.jp, and inutsuka@nagoya-u.jp}
\begin{abstract}
We perform a linear stability analysis of a dust layer in a turbulent gas disk.
Youdin (2011) investigated the secular gravitational instability of a dust
layer using hydrodynamic equations with a turbulent diffusion term.  We obtain
essentially the same result independently of Youdin (2011).  In the present
analysis, we restrict the area of interest to small dust particles, while
investigating the secular gravitational instability in a more rigorous manner.
We discuss the time evolution of the dust surface density distribution using a
stochastic model and derive the advection-diffusion equation.  The validity of
the analysis by Youdin (2011) is confirmed in the strong drag limit.  We
demonstrate quantitatively that the finite thickness of a dust layer weakens
the secular gravitational instability and that the density-dependent diffusion
coefficient changes the growth rate.  We apply the obtained results to the
turbulence driven by the shear instability and find that the secular
gravitational instability is faster than the radial drift when the gas density
is three times as large as that in the minimum-mass disk model. If the dust
particles are larger than chondrules, the secular gravitational instability
grows within the lifetime of a protoplanetary disk.
\end{abstract}

\section{Introduction \label{sec:intro}}
Planetesimals are solid objects that are assumed to exist in protoplanetary disks.  
It is thought that they had sizes larger than a kilometer in the solar nebular, which is large enough for gas drag forces to be insignificant compared to gravitational scattering.  The terrestrial planets and the cores of gas giants are considered to have been formed by the collisional accretion of planetesimals. However, the planetesimal formation process in protoplanetary disks remains controversial.

Small dust grains ranging in radii from a few microns to a few centimeters, grow by collisional sticking \citep[e.g.,][]{Dominik1997, Wurm1998}. However, the sticking force between dust aggregates having radii of larger than one centimeter is so weak that the growth of such aggregates may stall.
Dust collisions lead to fragmentation of dust aggregates because of the large collision velocity \citep[e.g.,][]{Blum2000}. Moreover, meter-sized dust aggregates are strongly affected by gas drag and drift radially on account of the headwind from the slower rotating gas. 
The timescale of the drift is approximately $100\, \mathrm{years}$ at 1 AU \citep{Adachi1976, Weidenschilling1977}.

If the gas flow is laminar, the dust aggregates settle toward the mid-plane, and the dust layer becomes very thin. Such a dense layer may be gravitationally unstable \citep{Safronov1972, Goldreich1973, Sekiya1983, Yamoto2004}. 
Dust-rich gas clumps are formed by the gravitational instability (GI) on the
dynamical timescale \citep{Sekiya1983, Yamoto2006}.  In the clump, dust
aggregates sediment to its center and a planetesimal forms \citep{Sekiya1983, Cuzzi2008}. 
The planetesimal formation timescale in the clump is longer than
the dynamical timescale, but its size is expected to be the same as the
classical estimate \citep{Sekiya1983}. 
Therefore, the gravitational instability model has been considered as a promising mechanism of planetesimal formation.

However, if the gas is turbulent, the dust aggregates are stirred up and the dust layer is not gravitationally unstable. Various mechanisms for driving turbulence have been proposed, including magneto-rotational instability \citep{Balbus1991, Sano2000} and shear instability \citep{Weidenschilling1980, Sekiya2000}. The resulting turbulence prevents the dust layer from settling, and the GI may not occur. 

One possible solution to this problem is the streaming instability model \citep{Youdin2005}.
When dust particles are moderately coupled to gas, they clump strongly enough to undergo gravitational collapse due to the streaming instability \citep{Johansen2007, Johansen2009}.
\cite{Johansen2007} found that large planetesimals, i.e., Ceres-sized planetesimals, can form directly by the streaming instability and subsequent GI.
Planetesimal formation by the streaming instability requires dust aggregates in the protoplanetary disk to become decimeter-sized particles.
However, the mechanism by which dust particles can grow to decimeter-size is unknown.
If dust particles are small and the drag force is strong, streaming turbulence develops \citep{Johansen2007b}. 
The condition for the dust clumping was investigate by the
numerical simulations in detail \citep{Bai2010, Bai2010a}. The
condition depends on the dust size distribution, the dust-to-gas mass ratio and
the gas pressure gradient \citep{Bai2010a}. The clumping due to the
streaming instability is feasible for the small radial gas pressure
gradient, the large size dust and the large dust surface
density.

\cite{Sekiya1998} investigated the dust density distribution in the turbulence driven by shear instability and found that the dust density is larger than the Roche density, the critical density for GI, when the dust-to-gas mass ratio is much larger than the solar abundance.
Thus, if dust density enhancement mechanisms exist, planetesimals can form due to GI.
\cite{Youdin2002} and \cite{Youdin2004} investigated the evolution of the dust distribution considering the radial drift of dust particles.
Since the inward mass flux decreases with decreasing radial distance from the Sun, dust particles tend to pile up, which may increase the dust-to-gas mass ratio.

The secular GI can also enhance the dust-to-gas mass ratio \citep{Ward1976, Coradini1981, Ward2000, Goodman2000, Michikoshi2010}.
If we consider the drag force due to gas, the dust layer is always secularly unstable, even when $Q>1$, where $Q$ is Toomre's $Q$ value \citep{Toomre1964}.
The growth timescale of the secular GI is longer than that of the GI without gas drag.
Previous studies adopted a simple hydrodynamic model.
However, this formulation may not be able to correctly handle the effects of the turbulence, and whether the secular GI can develop in turbulence remains unclear.
If strong turbulence exists, dust particles may be stirred, and the dust layer may not collapse.

\cite{Youdin2011} extended previous research efforts by considering the turbulent diffusion term. He performed a linear stability analysis using an isothermal hydrodynamic equation with a turbulent diffusion term. This equation can be applied to a wide range of parameter space.
He found that the secular GI can develop even when Toomre's $Q$ is greater than unity.
When the gas drag force is strong, the turbulent diffusion term is significant. The added diffusion term plays a crucial role. He used the $\alpha$ model for the turbulent diffusion \citep{Shakura1973}. Comparing the timescale of the secular GI with the disk lifetime and the radial drift time, he derived the upper limit of $\alpha$ value, which is approximately $10^{-8}$ to $10^{-3}$. Using the turbulence model, he obtained the threshold metallicity $Z$.
If dust particles cannot grow large, the secular GI may be more important than the streaming instability.

We have investigated the particle density evolution in turbulence and have
derived a similar result independently of \cite{Youdin2011}. 
In the present study, we restrict our investigation to small dust particles with a very strong drag force. Although the results of \cite{Youdin2011} are more general than the results of the present study, we derive time evolution equations for small dust particles in the turbulence in a more rigorous manner. \cite{Youdin2011} used an intuitive model for the basic equations of his study, which is a hydrodynamic model with a turbulent diffusion term. We verify the results of \cite{Youdin2011} for small dust particles. If the dust particles are small, the basic equation is described by a single advection-diffusion equation. In this case, the dispersion relation and eigenfunction can be calculated in a simple analytic form. Thus, the results of the present study are easier to understand. \cite{Youdin2011} used a simple turbulence model to calculate the threshold metallicity. 
We adopt the shear turbulence model given by \cite{Sekiya1998} and discuss the possibility and condition for the secular GI in shear turbulence.

The remainder of the present paper is organized as follows. In Section \ref{sec:basic}, we derive the time evolution equation for the dust density distribution for small particles from the Langevin equations. This formulation can handle the turbulent diffusion correctly. In Section \ref{sec:linear}, using the derived time evolution equation, we perform a linear stability analysis of the dust layer in turbulence. In Section \ref{sec:diskmodel}, we describe the shear turbulence model adopted herein. In Section \ref{sec:instability}, we calculate the growth timescale of the secular GI and discuss the possibility of the secular GI. In Section \ref{sec:conclusion}, we summarize the results.

\section{Time Evolution Equation \label{sec:basic}}
In this section, we discuss the density evolution of small dust particles in a turbulent disk.
Particles diffuse in phase space because of the random force from turbulence.
The general formula of the diffusion coefficient due to turbulence was given by \cite{Youdin2007}.
We consider an external force field and diffusion in the radial direction as well as the azimuthal direction, which is not considered in \cite{Youdin2007}.
The external force field does not change the diffusion coefficient but determines the drift term
if the force gradient is small.
The strict derivation of the time evolution equation of the surface density with the external force field in the general parameter will be investigated in a future study.
We assume a thin dust layer and neglect vertical motion of dust.

\subsection{Equation of Motion}
We investigate the dynamics of dust particles in turbulence using the stochastic differential equation. We consider very small dust aggregates, the stopping time $t_\mathrm{s}$ due to gas drag of which is very short, $t_\mathrm{s} \ll T_\mathrm{K}$, where $T_\mathrm{K}$ is the orbital period.

Using the local Cartesian coordinate system $(x,y)$, we measure the particle velocity $(v_x, v_y)$ relative to the local Keplerian shear: 
$v_x = dx/dt$, $v_y = dy/dt + (3/2) \Omega x$, where $\Omega$ is the Kepler frequency.

The equations of motion for dust particles are
\begin{equation}
\frac{d v_x}{ dt} = 2 \Omega v_y - \gamma (v_x - v_{\mathrm{g}x}) + f_x,
\label{eq:eomx}
\end{equation}
\begin{equation}
\frac{d v_y}{ dt} = - \frac{1}{2} \Omega v_x - \gamma (v_y - v_{\mathrm{g}y}) + f_y,
\label{eq:eomy}
\end{equation}
where $(f_x,f_y)$ is the external force that depends on the particle position $(x,y)$, and $(v_{\mathrm{g}x},v_{\mathrm{g}y})$ is the turbulence velocity.
The drag coefficient $\gamma=1/t_\mathrm{s}$ is constant.
The gas velocities are stochastic variables, and Equations (\ref{eq:eomx}) and (\ref{eq:eomy}) are stochastic differential equations and are referred to as Langevin equation in statistical physics.

We assume that the stopping time due to gas drag is shorter than the Kepler time and the timescale over which the turbulence velocity changes.  
Then, the dust velocity is approximated by the terminal velocity.
The terminal velocity is given by the force balance:
\begin{equation}
v_x = \frac{  v_{\mathrm{g}x} \gamma^2 + ( 2\Omega v_{\mathrm{g}y} +  f_x) \gamma + 2f_y \Omega}{\gamma^2 + \Omega^2},
\label{eq:s1}
\end{equation}
\begin{equation}
v_y = \frac{2 v_{\mathrm{g}y} \gamma^2 + (- \Omega v_{\mathrm{g}x} + 2f_y) \gamma -  f_x \Omega}{2(\gamma^2 + \Omega^2)}.
\label{eq:s2}
\end{equation}

\subsection{Turbulence Model}
We assume that the ensemble mean of turbulent gas velocity is the Keplerian velocity.
Thus, we obtain $\langle v_{\mathrm{g}x} \rangle = \langle v_{\mathrm{g}y} \rangle = 0$, where the angled bracket denotes an ensemble average.

We consider the turbulence velocity $v(t)$ to be a random time series.
The auto-correlation is the cross-correlation of the time series with itself $\phi(t) = \langle v(t_0) v(t_0 + t) \rangle$. If $v(t)$ is stationary, $\phi(t)$ does not depend on the time $t_0$. 
In general, $\phi(t)$ is a decreasing function of $t$. The typical decay time of $\phi(t)$ is the correlation time $t_\mathrm{c}$. If we assume the Kolmogorov spectrum, the auto-correlation is described by the following exponential function \citep{Youdin2007}:
\begin{equation}
\phi(t) = \sigma^2 \exp \left(-\frac{|t|}{t_\mathrm{c}} \right),
\label{eq:acf}
\end{equation}
where $\sigma$ is the velocity dispersion of turbulence.

We assume that the correlation time $t_\mathrm{c}$ is equal to the eddy turnover time $t_\mathrm{e}$.
The typical eddy turnover time is $t_\mathrm{e} \sim \Omega^{-1}$ \citep{Sekiya1998}. 
We define the gas diffusion coefficients as follows (See Appendix A):
\begin{equation}
D_{\mathrm{g}xx} = \sigma_{xx} t_\mathrm{e}^2,
\end{equation}
\begin{equation}
D_{\mathrm{g}yy} = \sigma_{yy} t_\mathrm{e}^2,
\end{equation}
\begin{equation}
D_{\mathrm{g}xy} = \sigma_{xy} t_\mathrm{e}^2,
\end{equation}
where $\sigma_{xx}$ and $\sigma_{yy}$ are the turbulence velocity dispersion,
and $\sigma_{xy}$ is the magnitude of the cross correlation.

\subsection{Advection-Diffusion Equation}
The change in the position of particles due to turbulence is assumed to be small.
Then, the time evolution equation for the surface density $\Sigma(x,y,t)$ is given by \citep[e.g.,][]{Binney2008}
\begin{equation}
\frac{\partial \Sigma}{\partial t} = - \frac{\partial }{\partial x} \left(V_x \Sigma \right)
- \frac{\partial }{\partial y} \left(V_y \Sigma  \right)
+ \frac{1}{2} \frac{\partial^2 }{\partial^2 x} \left(D_{xx} \Sigma  \right)
+ \frac{1}{2} \frac{\partial^2 }{\partial^2 y} \left(D_{yy} \Sigma  \right)
+ \frac{\partial^2 }{\partial x \partial y} \left(D_{xy} \Sigma  \right),
\label{eq:fp}
\end{equation}
where $V_x$ and $V_y$ are the advection velocities, and $D_{xx}$, $D_{yy}$, and $D_{xy}$ are the diffusion coefficients.

We assume that the gradient of the external force is small.
We obtain the following first-order coefficients, which are the drift terms (see Appendix \ref{sec:derivation}):
\begin{equation}
V_{x} = \lim_{\Delta t \to 0} \frac{\langle \Delta x \rangle}{\Delta t} = \frac{\gamma f_x  + 2 \Omega f_y }{\gamma^2 + \Omega^2},
\label{eq;coef_01_x}
\end{equation}
\begin{equation}
V_{y} =  \lim_{\Delta t \to 0} \frac{\langle \Delta y \rangle}{\Delta t} = \frac{- \Omega f_x + 2 \gamma f_y }{2(\gamma^2 + \Omega^2)} - \frac{3}{2} \Omega x.
\label{eq;coef_01_y}
\end{equation}
where $\Delta x=x-x_0$ and $\Delta y=y-y_0$ are the changes of $x$ and $y$ over the time interval $\Delta t = t-t_0 $, $t_0$ is the initial time, and $(x_0,y_0)$ is the initial position.
We assume that $\Delta t$ is smaller than the typical timescale $\Delta t< T$ but is larger than the eddy turnover time $\Delta t > t_\mathrm{e}$.
These velocities are equal to the terminal velocity of dust without turbulent stirring.

The second-order coefficients are as follows (see Appendix \ref{sec:derivation}):
\begin{equation}
D_{xx} = \lim_{\Delta t \to 0} \frac{1}{2} \frac{\langle \Delta x^2\rangle}{\Delta t} = \frac{\gamma^4 D_{\mathrm{g}xx} + 4 \Omega \gamma^3 D_{\mathrm{g}xy} + 4 \Omega^2 \gamma^2 D_{\mathrm{g}yy}}{(\gamma^2 + \Omega^2)^2},
\label{eq:difx}
\end{equation}
\begin{equation}
D_{yy} = \lim_{\Delta t \to 0}  \frac{1}{2} \frac{\langle \Delta y^2\rangle}{\Delta t} =\frac{\gamma^2 \Omega^2 D_{\mathrm{g}xx} - 4 \Omega \gamma^3 D_{\mathrm{g}xy} + 4 \gamma^4 D_{\mathrm{g}yy}}{4(\gamma^2 + \Omega^2)^2},
\label{eq:dify}
\end{equation}
\begin{equation}
D_{xy} = \lim_{\Delta t \to 0} \frac{1}{2} \frac{\langle \Delta x \Delta y \rangle}{\Delta t} =\frac{-\gamma^3 \Omega D_{\mathrm{g}xx} + 2 \gamma^2 ( \gamma^2 + \Omega^2) D_{\mathrm{g}xy} + 4 \Omega \gamma^3 D_{\mathrm{g}yy}}{2(\gamma^2 + \Omega^2)^2}.
\label{eq:difxy}
\end{equation}
These coefficients are the diffusion coefficients due to turbulent stirring.
The radial diffusion coefficient $D_{xx}$ is the same as the diffusion coefficient in \cite{Youdin2007}. 

If gas drag is strong, then dust particles are well coupled with the gas and the diffusion coefficient for the dust is the same as that for the gas $\gamma \gg \Omega$, $D_{xx} \simeq D_{\mathrm{g}xx}$, $D_{yy} \simeq D_{\mathrm{g}yy}$, $D_{xy} \simeq D_{\mathrm{g}xy}$.
We can use this approximation for small dust particles that are much smaller than $1 \mathrm{m}$.

When we consider the external force as the self-gravity of the dust layer, the external forces of the $x$ and $y$ components are
\begin{equation}
f_x = - \frac{\partial \phi}{\partial x},
\label{eq:forcex}
\end{equation}
\begin{equation}
f_y = - \frac{\partial \phi}{\partial y},
\label{eq:forcey}
\end{equation}
where $\phi$ is the gravitational potential.
The gravitational potential is given by the Poisson equation:
\begin{equation}
\left(\frac{\partial^2}{\partial x^2} + \frac{\partial^2}{\partial y^2} + \frac{\partial^2}{\partial z^2} \right) \phi = 4 \pi G \rho,
\label{eq:poisson}
\end{equation}
where $\rho$ is the dust density.

\subsection{Comparison with Youdin (2011)}
The time evolution equation derived here is the time evolution equation for the special case in \cite{Youdin2011}, in which the hydrodynamic equation was considered and an additional
diffusion term was added to the equation of continuity.
Since the gas drag is assumed to be strong, the Coriolis force is negligible.
The time evolution equations in \cite{Youdin2011} without the Coriolis force are
\begin{equation}
\frac{\partial \Sigma}{\partial t} + \frac{\partial (\Sigma v_x)}{\partial x} = D \frac{\partial^2 \Sigma}{\partial x^2},
\label{eq:eoc_youdin}
\end{equation}
\begin{equation}
\frac{\partial v_x}{\partial t} = - \gamma v_x  + \frac{c^2}{\Sigma} \frac{\partial \Sigma}{\partial x} +f_x,
\end{equation}
where $c$ is the velocity dispersion, and $D$ is the gas diffusion coefficient.

If the gas drag force is strong, then the mass flux is estimated in terms of the terminal velocity:
\begin{equation}
\Sigma v_x  = \frac{c^2}{\gamma } \frac{\partial \Sigma}{\partial x} +\frac{\Sigma f_x}{\gamma}.
\label{eq:term}
\end{equation}
Substituting Equation (\ref{eq:term}) into Equation (\ref{eq:eoc_youdin}), we obtain 
\begin{equation}
\frac{\partial \Sigma}{\partial t}  + \frac{\partial (\Sigma f_x / \gamma)}{\partial x}  = D_\mathrm{eff} \frac{\partial^2 \Sigma}{\partial x^2}
\end{equation}
where $D_\mathrm{eff}$ is the effective diffusion coefficient:
\begin{equation}
D_\mathrm{eff} = D + \frac{c^2}{\gamma}.
\end{equation}
The model of the velocity dispersion and the diffusion coefficient in \cite{Youdin2011} is as follows:
\begin{equation}
c = \frac{\sqrt{1+2 \tau_\mathrm{s}^2 + (5/4) \tau_\mathrm{s}^3}}{1+\tau_\mathrm{s}^2} (D_\mathrm{g} \Omega)^{1/2},
\end{equation}
\begin{equation}
D = \frac{1 + \tau_\mathrm{s} + 4 \tau_\mathrm{s}^2}{(1+\tau_\mathrm{s}^2)^2} D_\mathrm{g},
\end{equation}
where $\tau_\mathrm{s} = \Omega t_\mathrm{s}$, and $D_\mathrm{g}$ is the gas diffusion coefficient.
For the strong gas drag limit, the velocity dispersion and the diffusion coefficient are $c\simeq (D_\mathrm{g} \Omega)^{1/2}$ and $D \simeq D_\mathrm{g}$. 
Thus, we obtain
\begin{equation}
D_\mathrm{eff} = D_\mathrm{g}\left(1 + \Omega t_\mathrm{s}\right) \simeq D_\mathrm{g}.
\end{equation}
The effective diffusion coefficient is equal to the diffusion coefficient that we derived for the case in which $t_\mathrm{s} \ll t_\mathrm{e}$.

If we replace the advection term by the terminal velocity in the equation of continuity with a diffusion term, then the equation of \cite{Youdin2011} is identical to that of the present paper.
When the gas drag is strong, by solving the equation of motion, the velocity is approximated by the terminal velocity with the pressure gradient.
Since the pressure gradient term is small compared to the diffusion term in the strong drag limit, the terminal velocity is equal to that in the present paper.
The basic equation of the present paper is derived from the strict stochastic model.
Thus, the formulation in \cite{Youdin2011} is valid at least if the gas drag is strong.

If the gas drag is weak, the formulation of the present paper, i.e., the advection-diffusion equation, breaks down.
However, the hydrodynamic equation with a diffusion term for the weak drag limit has not yet been verified.
When the gas drag is weak, the relaxation of the velocity distribution is very slow.
In this case, the hydrodynamic picture may break down.
The time evolution equation for the weak drag limit is arguable.
Although we may qualitatively understand the physical nature of the instability using the hydrodynamic equation with a diffusion term, in order to discuss the physical nature of the instability quantitatively, we should investigate the time evolution equation carefully.
This issue will be discussed in detail in a future paper.

\section{Linear Stability Analysis\label{sec:linear}}
\subsection{Infinitesimally Thin Disk}
We consider only axisymmetric modes in order that a normal mode analysis can be performed.
We hereinafter omit the subscript $x$.
From Equations (\ref{eq:fp}) and (\ref{eq;coef_01_x}), the basic equations are 
\begin{equation}
\frac{\partial \Sigma}{\partial t} = - \frac{\partial }{\partial x} \left(V \Sigma \right)
+ \frac{\partial^2 }{\partial^2 x} \left(D \Sigma  \right)
\end{equation}
\begin{equation}
V = \frac{\gamma f}{\gamma^2 + \Omega^2} = \frac{f}{\gamma '},
\end{equation}
where $\gamma ' = \gamma + \Omega^2 / \gamma$ is the effective drag coefficient.
For the strong gas drag $\gamma \gg \Omega$, we obtain $\gamma ' \simeq \gamma$.

We perform the linear stability analysis on axisymmetric modes.
The unperturbed state is uniform in space and time.
We consider the perturbation of the form of $\exp(-i(\omega t - kx))$, where $k$ is the wave number and $\omega$ is the frequency.
The unperturbed and perturbed quantities are denoted by the subscripts `0' and `1', respectively.

Using the thin disk approximation, we can solve the Poisson equation \citep{Toomre1964},
\begin{equation}
\phi_1 = - \frac{2 \pi G \Sigma_1}{|k|}.
\end{equation}
The drift velocity of the $x$ component is 
\begin{equation}
V_{1} = \frac{- i k \phi_1 }{\gamma '}.
\end{equation}
The linear advection-diffusion equation is 
\begin{equation}
- i \omega \Sigma_1 =  - i k V_{1} \Sigma_0 - k^2 D \Sigma_1.
\end{equation}

Introducing the growth rate $\mu = \omega / i$, we obtain the dispersion relation:
\begin{equation}
\mu = -D k^2 + \frac{2 \pi |k| \Sigma_0 G}{\gamma '}
\label{eq:drel}
\end{equation}

If the wave number $k$ is less than the critical wave number
\begin{equation}
k_\mathrm{cr} = \frac{2 \pi \Sigma_0 G }{D \gamma '},
\end{equation}
the dust layer is unstable.
This feature is similar to the secular GI for the hydrodynamic equation model \citep{Michikoshi2010, Youdin2011}.
The dispersion relation indicates that the unstable mode always exists, regardless of the diffusion term. The turbulent diffusion makes the growth rate small, but it cannot stabilize all modes.
The diffusion term is effective for the short wavelength mode. However, the long wavelength mode is still unstable, even though the turbulent diffusion is included.

The maximum growth rate is 
\begin{equation}
\mu_\mathrm{max} = \frac{\pi^2 \Sigma_0^2 G^2 }{D \gamma '^2},
\end{equation}
and the wave number of the maximum growth rate is
\begin{equation}
k_\mathrm{max} = \frac{1}{2} k_\mathrm{cr} = \frac{\pi \Sigma_0 G }{D \gamma '}.
\end{equation}
The maximum growth rate is a decreasing function of the diffusion coefficient.
This maximum growth rate and the wave number are identical to Equation (56) in \cite{Youdin2011} with $D= D_\mathrm{g}$. This indicates that the analysis of \cite{Youdin2011} is valid when $t_\mathrm{s} \ll t_\mathrm{e}$.

\begin{figure}
 \begin{center}
  	\plotone{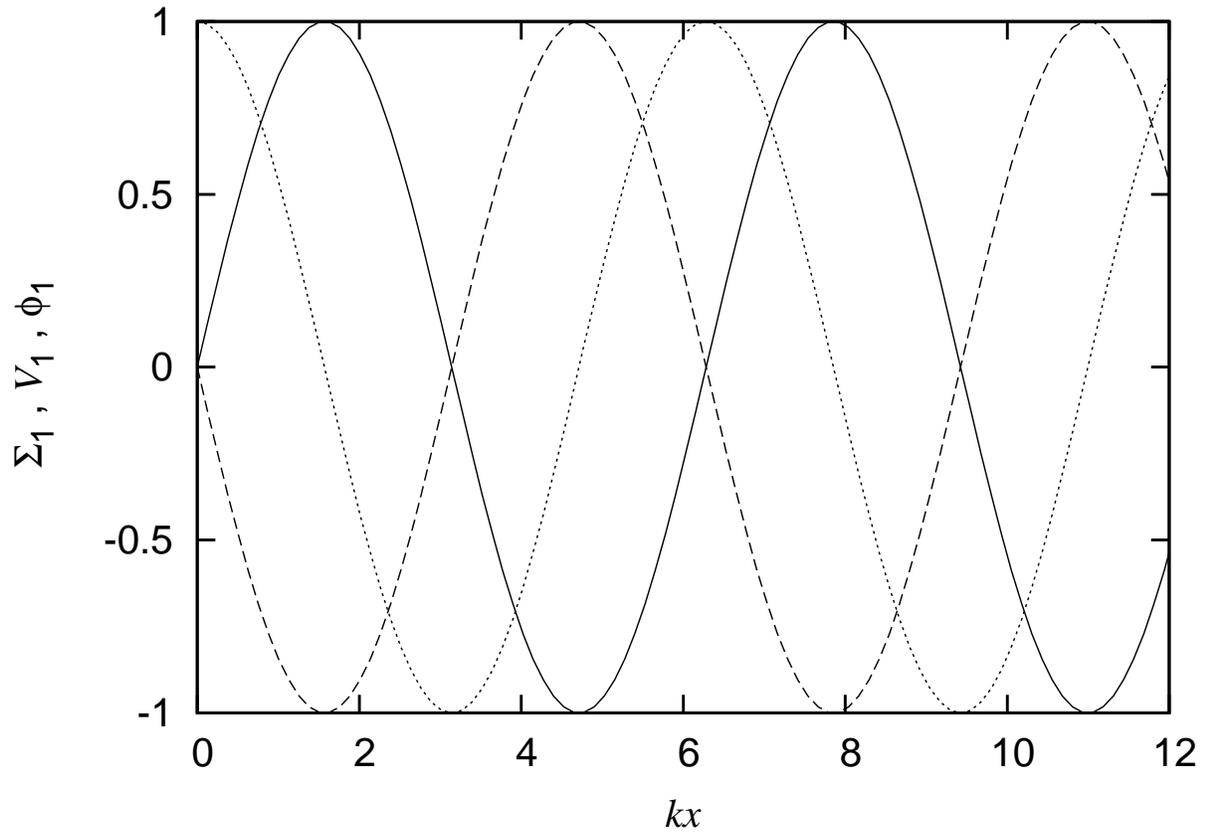}
 \end{center}
 \caption{Eigenfunctions of $\Sigma_1$ (\textit{solid curve}), $\phi_{1}$ (\textit{dashed curve}), and $V_{1}$ (\textit{dotted curve}). These eigenfunctions are normalized by amplitude. }
  \label{fig:eigen}
\end{figure}

The physical meaning of the secular GI can be understood in terms of the responses to the density perturbation. This is the diffusive support secular GI explained in Section 2.2 of \cite{Youdin2011}. The responses of the potential and the velocity to the density perturbation are shown in Figure \ref{fig:eigen}. We assume a sinusoidal density fluctuation. The gravitational potential $\phi_1$ is induced by the density fluctuation. Due to the resultant potential gradient, the matter moves to the local maximum point of the surface density at terminal velocity. Therefore, the density fluctuation increases monotonically.

The timescale of the secular GI and the stabilization by the diffusion can be understood as follows.
First, we neglect the effect of the diffusion. We suppose a wave with wavelength $L$.
The specific force at a point located at distance $L$ from an infinite straight line with line density $\mu$ is $\sim G \mu / L$.
We approximate the gravitational field from the wave as the gravitational field from an infinite straight line with line density $\mu \sim L \Sigma_1$.
From the force balance between the gravitational force and the drag force, the terminal velocity is given as $V_{1} \simeq G \Sigma_1 / \gamma '$.
The resulting mass flux is $\Sigma V_1 \simeq G\Sigma_1 \Sigma_0 / \gamma '$.
Therefore, the timescale of the increase in the surface density is $\Sigma_1 / (G\Sigma_1 \Sigma_0 / (L\gamma ')) \simeq L \gamma ' / G \Sigma_0$. This corresponds to the growth time calculated by Equation (\ref{eq:drel}) with $D=0$. This mode should be stabilized if the diffusion time $L^2/D$ is shorter than the timescale of the secular GI.
Thus, the critical length scale for the secular GI is $L_\mathrm{cr} \sim D \gamma '/G\Sigma_0$.
This estimation is consistent with the linear stability analysis.
The diffusion can smooth out only small-scale density fluctuation.
The long wavelength mode is not stabilized by the diffusion.

\subsection{Finite Thickness Disk}
For the sake of simplicity, the dust layer is assumed to be infinitesimally thin. However, in reality, the dust particles are stirred by the turbulence, and the dust layer has a finite thickness. We account for the effect of the finite thickness of the dust layer by introducing a softening term in the Poisson equation \citep{Vandervoort1970, Shu1984, Youdin2011}: $$ \phi_1 \simeq - \frac{2 \pi G \Sigma_1}{|k|(1+|k|h)},$$ where $h$ is the thickness of the dust layer. This expression indicates that the finite thickness reduces the effect of the self-gravity by a factor of $1/(1+|k|h)$.

\begin{figure}
 \begin{center}
  	\plotone{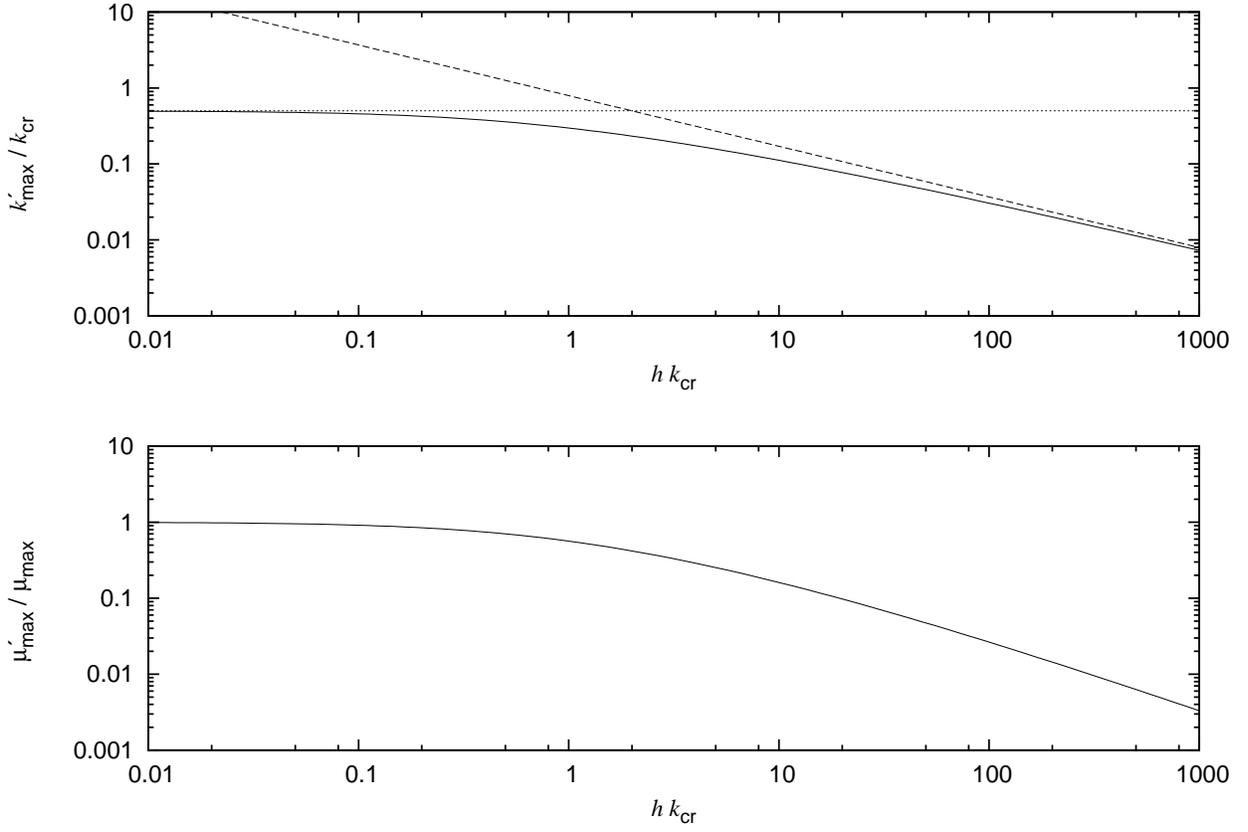}
 \end{center}
 \caption{
Wave number (\textit{top panel}) and growth rate (\textit{bottom panel}) for the most unstable mode as a function of $h k_\mathrm{cr}$. In the top panel, the dashed and dotted lines represent $k_\mathrm{cr}/(2^{1/3} (h k_\mathrm{cr})^{2/3})$ and $k_\mathrm{cr}/2$, which are asymptotic solutions of Equation (\ref{eq:wavenum_mu}). }
  \label{wavenum_mukmax}
\end{figure}

The dispersion relation for the finite thickness dust layer is
\begin{equation}
\mu = -D k^2 + \frac{2 \pi |k| \Sigma_0 G}{\gamma ' (1 + |k|h)}.
\end{equation}
The critical wave number for the finite-thickness disk is smaller than that for the infinitesimally thin disk: 
\begin{equation}
k_{\mathrm{cr}}'= \frac{2}{\sqrt{4 h k_\mathrm{cr} +1}+1} k_\mathrm{cr} \le k_\mathrm{cr}
\end{equation}
The wave number for the most unstable mode $k_\mathrm{max}'$ is given by $d \mu/dk =0$:
\begin{equation}
k_{\mathrm{max}}'^3 + 2 \frac{k_{\mathrm{max}}'^2}{h} + \frac{k_{\mathrm{max}}'}{h^2} - \frac{k_\mathrm{cr}}{2h^2}=0.
\label{eq:wavenum_mu}
\end{equation}
We calculate the leading term of the power series expansion with respect to $h$ and obtain the asymptotic solution of Equation (\ref{eq:wavenum_mu}).
When $h k_\mathrm{cr} \ll 1$, $k_{\mathrm{max}}'$ is approximated as $k_\mathrm{cr}/2$.
This limit corresponds to the infinitesimally thin approximation.
When $h k_\mathrm{cr} \gg 1$, $k_{\mathrm{max}}'$ is approximated as $k_\mathrm{cr}/(2^{1/3} (h k_\mathrm{cr})^{2/3})$.
The wave number for the most unstable mode is a decreasing function of thickness $h$.
As shown in Figure \ref{wavenum_mukmax}, $k_{\mathrm{max}}'$ is smaller than the most unstable wave number for the infinitesimally thin disk $k_{\mathrm{max}}'=k_\mathrm{cr}/2$.
We can prove that the real positive solution $k_{\mathrm{max}}'/k_\mathrm{cr}$ must be smaller than $1/2$ for any $h$.
The effect of the finite thickness is to elongate the most unstable wavelength.

The bottom panel of Figure \ref{wavenum_mukmax} shows the maximum growth rate of the secular GI $\mu_{\mathrm{max}}'$ as a function of the dust layer thickness $h$. 
When the dust layer is thick, the effect of the reduction in the self-gravity due to the dust thickness is significant.
Thus, the maximum growth rate is a decreasing function of the dust layer thickness.
If the dust layer is much thinner than the critical wavelength $\sim 1/k_\mathrm{cr}$, the effect of the thickness is negligible.
Therefore, the thickness factor $h k_\mathrm{cr}$ determines the effect of dust layer thickness.

\subsection{Density-dependent Diffusion Coefficient}
We have neglected the effect of back-reaction of the dust on the gas. If we consider the back-reaction, the turbulence velocity may change with the dust surface density. According to a numerical simulation by \cite{Johansen2009}, the particle random velocity in the dense region is small, and the resulting diffusion coefficient in the dense region is small. Thus, the diffusion coefficient $D$ may be a decreasing function of the surface density $\Sigma$. In general, $D$ is a function of $\Sigma$. Here, we investigate the density-dependent diffusion coefficient $D(\Sigma)$.

The diffusion coefficient is assumed to depend on the surface density:
\begin{equation}
D=D(\Sigma) \simeq D(\Sigma_0 + \Sigma _1 ) \simeq D_{0} + \frac{d D}{d \Sigma} \Sigma _1,
\end{equation}
where $D_{0} = D(\Sigma_0)$.
The diffusion term in the advection-diffusion equation is
\begin{equation}
\frac{\partial^2 (D \Sigma)}{\partial x^2} \simeq \left( D_{0} + \frac{d D}{d \Sigma} \Sigma _0 \right) \frac{\partial^2 \Sigma_{1}}{\partial x^2} .
\end{equation}
In other words, the term obtained from $d D/{d \Sigma}$ is added to the diffusion coefficient.
Thus, when we consider the density-dependent diffusion coefficient, we must replace the diffusion coefficient in Equation (\ref{eq:drel}) by
\begin{equation}
D_{0} + \frac{d D}{d \Sigma} \Sigma _0
= D_{0} \left(1 + \frac{d \log D}{d \log \Sigma} \right).
\end{equation}
Therefore, we obtain
\begin{equation}
\mu = -   D_{0}\left(1 - \beta  \right)k^2 + \frac{2 \pi |k| \Sigma_0 G}{\gamma '},
\end{equation}
where
\begin{equation}
\beta = - \frac{d \log D}{d \log \Sigma}.
\end{equation}
If we adopt the power-law model, we obtain $D\propto \Sigma ^{-\beta}$.

When $\beta>1$, any mode is unstable, although self-gravity is not considered.
Essentially, this is the same as the diffusion instability (viscous instability) that was discussed in explaining the ring structure of Saturn \citep{Lukkari1981, Lin1981}.
If $\beta>1$, $\Sigma D$ is a local minimum value when the density is a local maximum value.
Thus, the density at the local maximum point of the surface density increases
due to the resultant diffusion flux.
When $\beta=1$, the diffusion term vanishes. In this case, due to self-gravity, any mode is unstable.
When $\beta< 1$, the diffusion term stabilizes the short wavelength modes.
Thus, the maximum growth rate exists.
The maximum growth rate and the wave number are 
\begin{equation}
\mu_\mathrm{max} = \frac{\pi^2 \Sigma_0^2 G^2 }{(1-\beta)D_{0}\gamma '^2},
\end{equation}
\begin{equation}
k_\mathrm{max} = \frac{\pi \Sigma_0 G }{D_{0}(1-\beta)\gamma '}.
\end{equation}
The growth rate increases with $\beta$ because the effective diffusion coefficient decreases with $\beta$. 

In the present paper, we focus on the shear turbulence in the application.
The diffusion coefficient due to shear turbulence does not depend on $\Sigma$.
Thus, in the following, we will not discuss the secular GI for the density-dependent diffusion coefficient.
Note that this effect can be important in some types of turbulence.
In particular, when $\beta\sim 1$, the maximum growth rate depends sensitively on $\beta$.

\section{Shear Turbulence Model \label{sec:diskmodel}}
The linear stability analysis reveals that a small density fluctuation can grow in a turbulent disk 
regardless of the turbulence strength. We apply the secular GI to the turbulence due to the shear instability in a protoplanetary disk.

\subsection{Minimum-mass Disk Model}
The surface density is estimated based on the minimum-mass disk model \citep{Hayashi1981, Hayashi1985}.
The surface densities of gas and dust are
\begin{equation}
\Sigma_\mathrm{g} = 1.7 \times 10^3 f_\mathrm{g} \left(\frac{a}{\mathrm {AU}} \right)^{-3/2} \mathrm{g}/\mathrm{cm}^2,
\end{equation}
\begin{equation}
\Sigma_\mathrm{d} = 7.1 f_\mathrm{d} \left(\frac{a}{\mathrm {AU}} \right)^{-3/2} \mathrm{g}/\mathrm{cm}^2,
\end{equation}
where $a$ is the distance from the Sun, and $f_\mathrm{d}$ and $f_\mathrm{g}$ are enhancement factors of the surface densities of gas and dust.
In the minimum-mass disk model, we assume $f_\mathrm{d} / f_\mathrm{g} = 4.2$ outside $2.7\, \mathrm{AU}$ because $\mathrm{H}_2\mathrm{O}$ condenses.
The gas temperature $T$, the gas density $\rho_\mathrm{g}$ on the mid-plane, the sound velocity $c_\mathrm{s}$, and the pressure gradient $\eta$ are obtained as follows: 
\begin{equation}
T=280 \left(\frac{a}{\mathrm {AU}}\right)^{-1/2} \mathrm{K},
\end{equation}
\begin{equation}
\rho_{\mathrm{g}}=1.4 \times 10^{-9} f_\mathrm{g} \left(\frac{a}{\mathrm {AU}}\right)^{-11/4} \mathrm{g}/ \mathrm{cm}^3,
\end{equation}
\begin{equation}
c_\mathrm{s} = 1.1 \times 10^5 \left(\frac{a}{1\mathrm{AU}} \right)^{-1/4}\, \mathrm{cm}/\mathrm{s},
\end{equation}
\begin{equation}
\eta = 1.81 \times 10^{-3} \left(\frac{a}{\mathrm{AU}}\right)^{1/2}.
\end{equation}

\subsection{Structure of the Dust Layer}
The dust particles are assumed to be very small.
In this case, the gas drag force is strong.
As dust settles toward the mid-plane, shear instability is inevitable \citep{Sekiya2000, Sekiya2001, Ishitsu2002, Ishitsu2003, Michikoshi2006}.
The resultant turbulence prevents dust from settling further.
Consequently the dust layer attains a quasi-equilibrium state \citep{Sekiya1998}.

The stability of the shear flow is determined by the Richardson number $J$:
\begin{equation}
J=-\frac{g}{\rho}\frac{\partial \rho / \partial z}{(\partial v / \partial z)^2},
\end{equation}
where $g$ is the vertical gravitational acceleration. 
If $J$ is less than the critical value $J_\mathrm{cr}$, the shear flow becomes turbulent.
The linear stability analysis reveals that the critical value is $J_\mathrm{cr}=0.25$ \citep{Chandrasekhar1961}.
Numerical simulation without the Coriolis force confirmed the critical Richardson number to be approximately $0.25$ \citep{Barranco2009}.
Numerical simulations that include the effect of rotation indicate that $J_\mathrm{cr}\simeq0.1$ \citep{Chiang2008}. 
In the case of small dust, the critical Richardson number is
$J_\mathrm{cr} \simeq 0.2$ for the solar metallicity disk, which increases with
metallicity \citep{Lee2010}.
Therefore, the critical Richardson number is $J_\mathrm{cr} = 0.1 \-- 0.25$.
When we estimate various quantities, we use $J_\mathrm{cr} = 0.1$, which is an optimistic assumption.

\cite{Sekiya1998} argued that the dust density distribution is characterized by assuming that the Richardson number $J(z)$ is equal to the critical Richardson number $J_\mathrm{cr}$ everywhere.
This assumption was explored by the numerical simulations \citep{Barranco2009, Lee2010a, Lee2010}.
The Richardson number of the dust layer in a protoplanetary disk is \citep{Sekiya1998},
\begin{equation}
J(z)=-\left(z + \frac{4 \pi G}{\Omega^2} \int_0^z(\rho_\mathrm{g}+\rho_\mathrm{d}(z))dz \right) \frac{(\rho_\mathrm{g}+\rho_\mathrm{d}(z))^3}{(\eta a \rho_\mathrm{g})^2 \partial \rho_\mathrm{d} / \partial z},
\label{eq:richa}
\end{equation}
where $\rho_\mathrm{d}(z)$ is the dust density.
Solving $J(z)=J_\mathrm{cr}$, we can calculate the dust density distribution.

The thickness of the dust layer $h_\mathrm{d}$ is \citep{Sekiya1998}
\begin{equation}
h_\mathrm{d} = \sqrt{J_\mathrm{cr}} \eta a \sqrt{1-R^2},
\label{eq:thickness}
\end{equation} 
where $R$ is the ratio of the gas density to the total density at the mid-plane:
\begin{equation}
R = \frac{\rho_\mathrm{g}}{\rho_\mathrm{g} + \rho_\mathrm{d}(0)}.
\end{equation} 
The ratio $R$ is obtained as the solution of the following equation \citep{Sekiya1998}:
\begin{equation}
\frac{S}{1+q} =  \log \left(\frac{1+\sqrt{1-((R+q)/(1+q))^2}}{(R+q)/(1+q)} \right) -\sqrt{1-((R+q)/(1+q))^2},
\label{eq:den}
\end{equation} 
where $S= \Sigma_\mathrm{d}/(2\sqrt{J_\mathrm{cr}} \eta a \rho_\mathrm{g})$, and $q=4 \pi G \rho_\mathrm{g}/\Omega^2$ .
The parameters $S$ and $q$ in the minimum-mass disk model are
\begin{equation}
S= 0.30 \frac{f_\mathrm{d}}{f_\mathrm{g}} \left(\frac{a}{\mathrm{AU}}\right)^{-1/4} \left(\frac{J_\mathrm{cr}}{0.1}\right)^{-1/2},
\label{eq:S_est}
\end{equation}
\begin{equation}
q= 0.030 f_\mathrm{g} \left(\frac{a}{\mathrm {AU}}\right)^{1/4}. 
\label{eq:q_est}
\end{equation}

When $S=0.3$ and $q=0.03$, we obtain $R=0.60$, and thus $\rho_\mathrm{d}/\rho_\mathrm{g} \simeq 0.68$.
Since the dust density is much smaller than the Roche density
\citep{Sekiya1983}, the dynamical GI does not occur.
\cite{Sekiya1998} reported that the dust density can be larger than the Roche density if the dust-to-gas mass ratio is much larger than the solar abundance, $f_\mathrm{d}/f_\mathrm{g} \sim 20$, 
for $T=280\, \mathrm{K}$ at $1\, \mathrm{AU}$.
The critical dust-to-gas mass ratio depends on the disk parameters \citep{Youdin2002}. If 
the disk temperature is $100 \-- 170 \, \mathrm{K}$ at $1\, \mathrm{AU}$, the 
critical dust-to-gas mass ratio is about $2 \-- 10$.
In the present paper, we assume that the dust-to-gas mass ratio is the solar abundance $f_\mathrm{d}/f_\mathrm{g} \sim 1$ and investigate the possibility of the secular GI.

\subsection{Diffusion Coefficient}
Setting the mass flux due to the sedimentation velocity equal to the diffusion flux, we have \citep{Cuzzi1993, Sekiya1998},
\begin{equation}
-\Omega^2 t_\mathrm{s} z = \frac{D}{\rho_\mathrm{d}} \frac{\partial \rho_\mathrm{d} }{\partial z}
\label{eq:dif_settle}
\end{equation}
From Equations (\ref{eq:richa}) and (\ref{eq:dif_settle}), we obtain the diffusion coefficient $D$, which depends on $z$,
\begin{equation}
D(z) = J_\mathrm{cr} t_{\mathrm{s}} \eta^2 v_\mathrm{K}^2 \frac{\rho_\mathrm{g}^2 \rho_\mathrm{d}(z) }{(\rho_\mathrm{g}+\rho_\mathrm{d}(z))^3}F_\mathrm{g}^{-1},
\end{equation}
where $F_\mathrm{g}$ is the effect of the self gravity and has the following upper limit:
\begin{equation}
F_\mathrm{g} = 1 + \frac{q}{z} \int_0^z\left(1+\frac{\rho_\mathrm{d}}{\rho_\mathrm{g}} \right)dz  < 1 + q \left(1+\frac{\rho_\mathrm{d}(0)}{\rho_\mathrm{g}} \right) = 1 + \frac{q}{R}.
\end{equation}
Since $q/R \ll 1$ is satisfied in the standard model, we assume that $ F_\mathrm{SG} \simeq 1$.
Accordingly, the diffusion coefficient is 
\begin{equation}
D(z) \simeq J_\mathrm{cr} t_{\mathrm{s}} \eta^2 v_\mathrm{K}^2 \frac{\rho_\mathrm{g}^2 \rho_\mathrm{d}(z) }{(\rho_\mathrm{g}+\rho_\mathrm{d}(z))^3}.
\end{equation}

If $\rho_\mathrm{d}(0)/\rho_\mathrm{g}>1/2$, $D(z)$ has a maximum value when $\rho_\mathrm{d}(z)/\rho_\mathrm{g}=1/2$, then:
\begin{equation}
D_\mathrm{max} = \frac{4}{27} J_\mathrm{cr}t_{\mathrm{s}} \eta^2 v_\mathrm{K}^2.
\end{equation}
We assume the typical diffusion coefficient in the dust layer to be equal to half of $D_\mathrm{max}$:
\begin{equation}
D = \frac{2}{27} J_\mathrm{cr} t_{\mathrm{s}} \eta^2 v_\mathrm{K}^2.
\end{equation}
This is the vertical diffusion coefficient of the turbulence.
The vertical diffusion coefficient is assumed to be equal to the radial diffusion coefficient.

\cite{Youdin2011} used $D \sim t_\mathrm{s} \eta^2 V_\mathrm{K}^2$ (Equation (68) in \cite{Youdin2011}). 
The diffusion model used herein has the same dependence as the model of \cite{Youdin2011}, except for $J_\mathrm{cr}$, and the coefficient of the diffusion model used herein is much smaller.

\section{Secular Gravitational Instability of a Dust Layer\label{sec:instability}}
\subsection{Timescale and Wavelength}
In this section, the growth time of the secular GI in a protoplanetary disk with shear turbulence is estimated.
First, we examine the dust layer thickness.
We can calculate the thickness factor from Equation (\ref{eq:thickness}) as follows:
\begin{equation}
 h k_\mathrm{cr} = \sqrt{J_\mathrm{cr}} \eta a k_\mathrm{cr} \sqrt{1-R^2} < \sqrt{J_\mathrm{cr}} \eta a k_\mathrm{cr} = 0.12 f_\mathrm{d}\left(\frac{J_\mathrm{cr}}{0.1}\right)^{-1/2}.
\end{equation}
The thickness factor does not depend on the dust particle radius and the distance from the sun because they are canceled out. Since the thickness factor is less than unity in the standard model, we adopt the infinitesimally thin disk approximation in calculating the growth time of the secular GI.

The growth time of the secular GI in the thin disk approximation is
\begin{equation}
 t_\mathrm{GI} = \frac{D \gamma '^2}{\pi^2 G^2 \Sigma_\mathrm{d}^2} = \frac{2 \eta^2 v_\mathrm{K}^2 J_\mathrm{cr}}{27 \pi^2 G^2 \Sigma_\mathrm{d}^2 t_\mathrm{s}} .
\end{equation}

There are two models of gas drag: the Stokes drag and the Epstein drag.
If the dust particle radius is less than the mean free path of gas, the Epstein drag is appropriate \citep[e.g.,][]{Epstein1924, Adachi1976}.
On the other hand, if the dust particle radius is larger than the mean free path, the Stokes drag is appropriate \citep[e.g.,][]{Stokes1851, Adachi1976}.
The mean free path $l$ is \citep[e.g.,][]{Adachi1976, Nakagawa1986}
\begin{equation}
l \simeq f_\mathrm{g}^{-1} \left(\frac{a}{1 \mathrm{AU}}\right)^{11/4} \mathrm{cm}.
\end{equation}
Thus, when $a < a_\mathrm{cr}$ , the Stokes drag should be used, where
\begin{equation}
a_\mathrm{cr} = f_\mathrm{g}^\mathrm{4/11} \left(\frac{r_\mathrm{p}}{1\mathrm{cm}}\right)^{4/11} \mathrm{AU},
\label{eq:acr}
\end{equation}
and $a_\mathrm{cr}$ is determined by $l=r_\mathrm{p}$.
The stopping time is
\begin{equation}
\displaystyle{
 t_\mathrm{s} = \left\{ \begin{array}{ll}
 \displaystyle{\frac{2 \rho_\mathrm{p} r_\mathrm{p}^2}{3 c_\mathrm{s} l \rho_\mathrm{g}}} & (a<a_\mathrm{cr}) \\
\displaystyle{
\frac{\rho_\mathrm{p} r_\mathrm{p}}{c_\mathrm{s} \rho_\mathrm{g}}} & (a>a_\mathrm{cr}) 
  \end{array} \right.,
}
\end{equation}
where $\rho_\mathrm{p}$ is the bulk density of the dust particles, and $r_\mathrm{p}$ is the radius of the dust.
The growth time of the secular GI is
\begin{equation}
 t_\mathrm{GI} = \left\{ \begin{array}{ll}
\displaystyle{
 2.4\times10^7
 \frac{f_\mathrm{g}}{f_\mathrm{d}^2}
 \left(\frac{J_\mathrm{cr}}{0.1}\right)
 \left(\frac{\rho_\mathrm{p}}{3 \mathrm{g}/\mathrm{cm}^3}\right)^{-1}
 \left(\frac{r_\mathrm{p}}{1 \mathrm{mm}}\right)^{-2}
 \left(\frac{a}{1 \mathrm{AU}}\right)^{11/4}
}
\mathrm{years} & (a<a_\mathrm{cr}) \\
\displaystyle{
1.6\times10^6 
 \frac{f_\mathrm{g}}{f_\mathrm{d}^2}
 \left(\frac{J_\mathrm{cr}}{0.1}\right)
 \left(\frac{\rho_\mathrm{p}}{3 \mathrm{g}/\mathrm{cm}^3}\right)^{-1}
 \left(\frac{r_\mathrm{p}}{1 \mathrm{mm}}\right)^{-1}
}
\mathrm{years} & (a>a_\mathrm{cr}) 
  \end{array} \right. .
\end{equation}
The growth time for $a>a_\mathrm{cr}$ does not depend on the distance from the Sun $a$. 

The critical wavelength of the secular GI is
\begin{equation}
 \lambda_\mathrm{cr} = \frac{2\pi}{k_\mathrm{cr}} = \frac{D \gamma '}{\Sigma_\mathrm{d} G} = \frac{2 J_\mathrm{cr} \eta^2 v_\mathrm{K}^2}{27 \Sigma_\mathrm{d} G}  = 3.0 \times 10^{-2} f_\mathrm{d}^{-1} 
 \left(\frac{J_\mathrm{cr}}{0.1}\right)
 \left(\frac{a}{\mathrm{AU}}\right)^{3/2} \mathrm{AU} .
\label{eq:length}
\end{equation}
This does not depend on the radius of the dust particle.

\subsection{Instability Conditions}
\subsubsection{Comparison with the Radial Drift}
If the radial drift of dust particles due to gas drag is faster than the secular GI, the secular GI is inefficient. Instead, the particle pileup mechanism due to the radial drift enhances the surface density of dust \citep{Youdin2002, Youdin2004}. Here, we compare the timescales of the secular GI and the radial drift.

\cite{Youdin2011} examined the same condition. Comparing the timescale of the secular GI with the radial drift, he derived the critical turbulence alpha value $\alpha_\mathrm{max}$ (Equation (53) in \cite{Youdin2011}). He applied $\alpha_\mathrm{max}$ to the turbulence model $D=t_\mathrm{s} \eta^2 v_\mathrm{K}^2$ and calculated the metallicity thresholds.
In this section, we perform essentially the same analysis.
However, we assume the turbulence model reported by \cite{Sekiya1998}.

According to the linear stability analysis, the streaming instability is faster than the radial drift \citep{Youdin2005}. In the present paper, we consider small dust particles.
The wavelength of the streaming instability is very short for small dust particles.
The numerical simulation of tightly coupled dust aggregates suggest that the concentration caused by the streaming instability is not very effective \citep{Johansen2007}.
The clumps are small and short-lived. Thus, in the case of small dust particles, the streaming instability contributes to the turbulence, but may not promote planetesimal formation. 
In the present paper, we ignore the streaming instability and focus on the radial drift.

The radial drift velocity of dust particles is \citep{Nakagawa1986}
\begin{equation}
v_r = \frac{2 R^2 t_\mathrm{s} \Omega}{1+R^2 t_\mathrm{s}^2 \Omega^2} \eta v_\mathrm{K}.
\end{equation}
If the gas drag is strong $R t_\mathrm{s} \Omega \ll 1$, the radial velocity is approximated as
\begin{equation}
v_r \simeq 2 R^2 t_\mathrm{s} \Omega \eta v_\mathrm{K}.
\end{equation}
The radial drift time of dust particles is
\begin{equation}
t_{r} \equiv \frac{a}{v_r} = \frac{1 }{2 \eta R^2 \Omega^2 t_\mathrm{s} }.
\end{equation}
Thus, the ratio of the timescale of the secular GI to the radial drift is
\begin{equation}
\frac{t_\mathrm{GI}}{t_{r}} = \frac{4 J_\mathrm{cr} \eta^3 R^2 v_\mathrm{K}^2 \Omega^2}{27 \pi^2 G^2 \Sigma_\mathrm{d}^2}
= 5.0 f_\mathrm{d}^{-2}\left(\frac{J_\mathrm{cr}}{0.1}\right)
 \left(\frac{R}{0.6}\right)^{2}
 \left(\frac{a}{1 \mathrm{AU}}\right)^{5/2}.
\end{equation}
This ratio does not depend on the dust particle radius. As the dust surface density increases or the pressure gradient $\eta$ decreases, the secular GI becomes faster. The ratio $R$ is positive and smaller than unity. 

\begin{figure}
 \begin{center}
  	\plotone{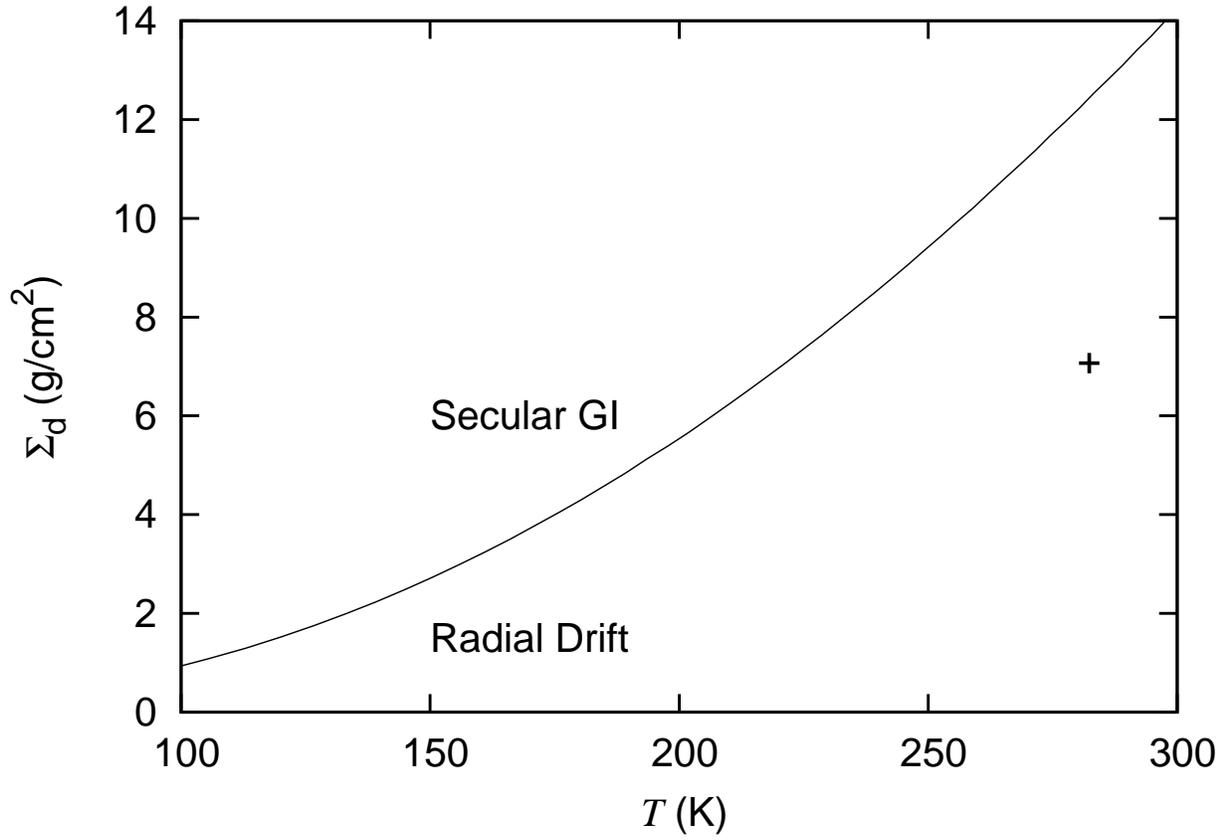}
 \end{center}
 \caption{Critical surface density for the secular GI at 1 $\mathrm{AU}$ as a function of the temperature $T$ for $f_\mathrm{g}=1$.
If the surface density is larger than the critical value, the secular GI is faster than the radial drift. The plus symbol denotes the standard minimum-mass disk model.}
  \label{fig:eta-sigma}
\end{figure}

We examine the temperature dependence. Since $\eta$ and $R$ depend on $T$,
the ratio $t_\mathrm{GI} / t_{r}$ depends on $T$ and $\Sigma_\mathrm{d}$.
Figure \ref{fig:eta-sigma} shows the condition for the secular GI in the $T$--$\Sigma_\mathrm{d}$ plane at 1 AU. We numerically calculate $R$ from Eq. (\ref{eq:den}).
In the case of the standard model, the radial drift is faster than the secular GI.
If the surface density is large or the temperature is low, the secular GI can occur.
In the case of $f_\mathrm{d}=1$, for the secular GI, the temperature must be lower than 220 K.
As the temperature decreases, the pressure gradient $\eta$ decreases.
Thus, the vertical shear and the resultant shear instability weaken.
In the case of $T=280\mathrm{K}$, if $f_\mathrm{d} > 1.7$, the secular GI is faster than the radial drift. In general, the realistic surface density can be larger than that of the minimum-mass disk model. We investigate the case in which the surface density is larger than that of the minimum-mass disk model, i.e., $f_\mathrm{g}, f_\mathrm{d} >1$.

\begin{figure}
 \begin{center}
  	\plotone{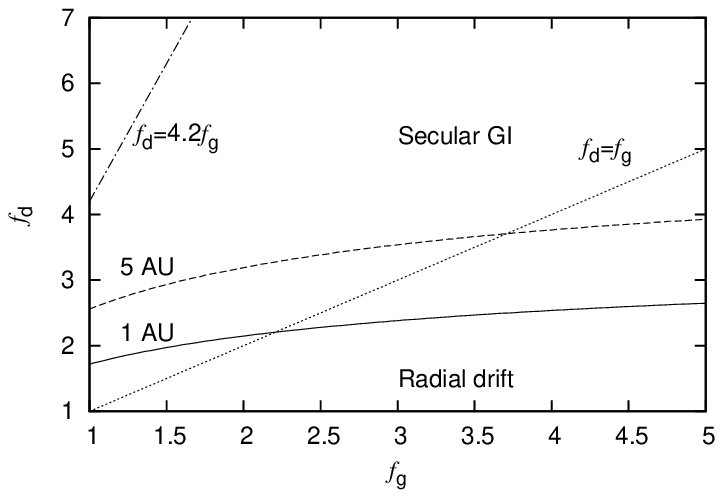}
 \end{center}
 \caption{Condition for the secular GI in the $f_\mathrm{g}$--$f_\mathrm{d}$ plane.
The solid curve denotes the critical $f_\mathrm{d,cr}$ value as a function of $f_\mathrm{g}$ for 1 AU. 
When $f_\mathrm{d} > f_\mathrm{d,cr}$, the secular GI is faster than the radial drift.
The dashed curve denotes the critical $f_\mathrm{d,cr}$ value for 5 AU.
The dotted line denotes the case in which $f_\mathrm{d} = f_\mathrm{g}$, which is the standard dust-to-gas mass ratio inside the snow line.
The dot-dashed line denotes the case in which $f_\mathrm{d} = 4.2 f_\mathrm{g}$, which is the standard dust-to-gas mass ratio outside the snow line.
}
  \label{fig:fd-fg}
\end{figure}

We calculate the condition in which the secular GI is faster than the radial drift in the $f_\mathrm{d}$--$f_\mathrm{g}$ plane. We define the critical value $f_\mathrm{d,cr}=f_\mathrm{d}$ at which $t_\mathrm{GI} / t_{r} =1$. When $f_\mathrm{d} > f_\mathrm{d,cr}(f_\mathrm{g},J_\mathrm{cr},a)$, the secular GI is faster than the radial drift.
Figure \ref{fig:fd-fg} shows that $f_\mathrm{d,cr}$ is an increasing function of $f_\mathrm{g}$ at $a = 1 \mathrm{AU}$, where we adopt $J_\mathrm{cr}=0.1$.
If $f_\mathrm{g}<2.2$, $f_\mathrm{d,cr}$ for 1 AU is larger than $f_\mathrm{g}$. 
Thus, the enhancement of the dust-to-gas mass ratio is necessary, so that the secular GI is faster than the radial drift for $f_\mathrm{g}<2.2$ at 1 AU.
On the other hand, for $f_\mathrm{g} > 2.2$, although the dust-to-gas mass ratio is the standard value $f_\mathrm{d}/f_\mathrm{g}=1$, the secular GI can be faster than the radial drift.
Therefore, if the protoplanetary disk is more massive than the minimum-mass disk, the secular GI can be effective at 1 AU, although the dust-to-gas mass ratio is the standard value.
Since $\mathrm{H}_2\mathrm{O}$ condenses at 5 AU, the standard dust-to-gas mass ratio is 4.2.
Thus, $f_\mathrm{d}$ is larger than $f_\mathrm{d,cr}$, although $f_\mathrm{g}=1.0$.

\begin{figure}
 \begin{center}
  	\plotone{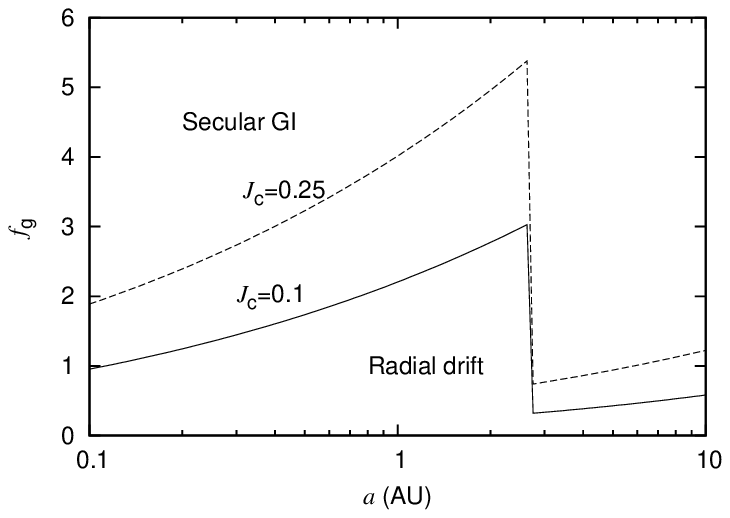}
 \end{center}
 \caption{
Critical gas enhancement factor $f_\mathrm{g,cr}$ as a function of the distance from the Sun $a$ for $J_\mathrm{cr}=0.1$ (solid curve) and that for $J_\mathrm{cr}=0.25$ (dashed curve).
 If $f_\mathrm{g}$ is larger than the critical $f_\mathrm{g,cr}$, the secular GI is faster for the standard dust-to-gas mass ratio.
For $a<2.7 \mathrm{AU}$, we calculate $f_\mathrm{g,cr}$ from $f_\mathrm{d,cr} / f_\mathrm{g,cr} = 1$.
For $a>2.7 \mathrm{AU}$, we calculate $f_\mathrm{g,cr}$ from $f_\mathrm{d,cr} / f_\mathrm{g,cr} = 4.2$ because $\mathrm{H}_2\mathrm{O}$ condenses.
  \label{fig:r_fgcrit}
 }
\end{figure}

Next, assuming the standard dust-to-gas mass ratio, we examine the condition in which the secular GI is faster than the radial drift in the $a$--$f_\mathrm{g}$ plane.
We define the critical value $f_\mathrm{g,cr}(J_\mathrm{cr},a)$ as $f_\mathrm{d,cr}(f_\mathrm{g,cr},J_\mathrm{cr},a) / f_\mathrm{g,cr} = 1$ for $a< 2.7 \mathrm{AU}$ and $f_\mathrm{d,cr}(f_\mathrm{g,cr},J_\mathrm{cr},a) / f_\mathrm{g,cr} = 4.2$ for $a> 2.7 \mathrm{AU}$. If $f_\mathrm{g} > f_\mathrm{g,cr}$, the secular GI is faster than the radial drift with the standard dust-to-gas mass ratio.
Figure \ref{fig:r_fgcrit} shows the critical $f_\mathrm{g,cr}$ as a function of the distance from the Sun $a$ for $J_\mathrm{cr}=0.1$ and $0.25$.
The critical value $f_\mathrm{g,cr}$ has a maximum value at $a=2.7 \mathrm{AU}$.
Therefore, for the case in which $J_\mathrm{cr}=0.1$, if $f_\mathrm{g}>3$, the secular GI is faster than the radial drift in the entire disk. 
When $J_\mathrm{cr} =0.25$, the turbulent diffusion is stronger than that for $J_\mathrm{cr} =0.1$.
Then, a more massive disk, such that $f_\mathrm{g}>5.5$, is necessary for the secular GI.

\subsubsection{Comparison with the Disk Lifetime}
\begin{figure}
 \begin{center}
  	\plotone{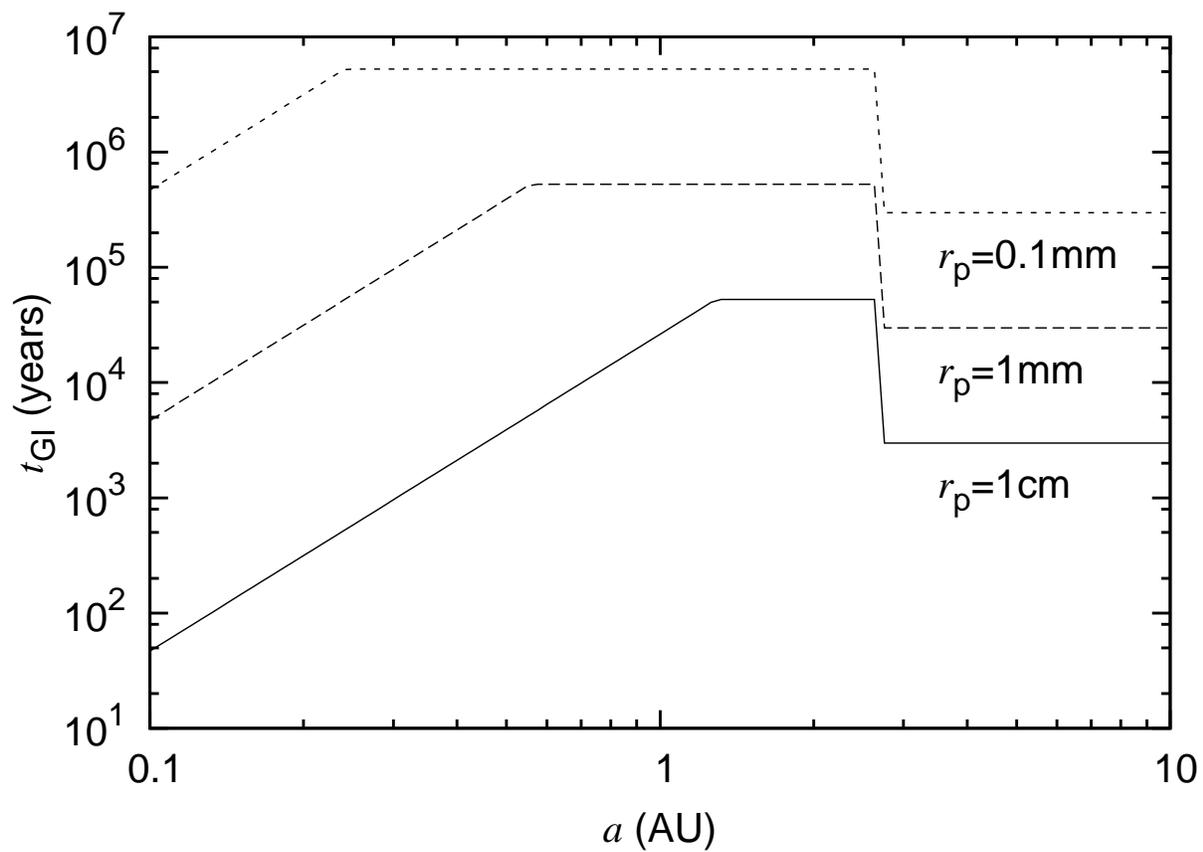}
 \end{center}
 \caption{Timescales of the secular GI for $r_\mathrm{p} = 1 \mathrm{cm}$ (solid curve), $r_\mathrm{p}=1 \mathrm{mm}$ (dashed curve) and $r_\mathrm{p}=0.1 \mathrm{mm}$ (short dashed curve). We assume that $f_\mathrm{g}=3$ and $J_\mathrm{cr}=0.1$.}
  \label{fig:r-tgi}
\end{figure}

The secular GI should occur within the disk lifetime.
The disk lifetime is typically $10^6 - 10^7$ years. Figure \ref{fig:r-tgi} shows the timescale of the secular GI. We adopt $f_\mathrm{g}=3$ and $J_\mathrm{cr}=0.1$. In the inner disk, the Stokes drag model is appropriate because the mean free path of the gas is less than the dust particle radius.
The timescale of the secular GI is proportional to $a^{11/4}$ in the inner disk.
On the other hand, in the outer disk, the Epstein drag model is appropriate because the mean free path of the gas is longer than the dust particle radius.
The timescale of the secular GI does not depend on the distance from the Sun $a$.
The discontinuity at $a=2.7 \mathrm{AU}$ corresponds to the snow line.
Since the surface density of dust is large outside the snow line, the secular GI is rapid.
Since the drag force from gas is weaker for larger dust particles, the timescale of the secular GI is faster for larger dust particles. The typical chondrule radius is $1\, \mathrm{mm}$. Thus, the growth time of the secular GI is $5.3 \times 10^{5}\, \mathrm{years}$.
The secular GI can grow within the disk lifetime.
If dust aggregates grow to $1\, \mathrm{cm}$, the growth time is $5.3 \times 10^{4}\, \mathrm{years}$, which is very rapid compared to the disk lifetime.

In Figure \ref{fig:r-tgi}, the growth time has a maximum value at the snow line $a=2.7\, \mathrm{AU}$.
From Equation (\ref{eq:acr}), the Epstein drag law is appropriate at $a=2.7\, \mathrm{AU}$ for
\begin{equation}
r_\mathrm{p} < 15.4 f_\mathrm{g}^{-1} \mathrm{cm}.
\end{equation}
Since we assume small dust particles, such as 1 $\mathrm{mm}$, this condition is satisfied. Therefore, the maximum growth time of the secular GI in the protoplanetary disk is evaluated in terms of Epstein drag.

The condition whereby the secular GI grows within the disk life time $t_\mathrm{disk}$ is given as follows:
\begin{equation}
r_\mathrm{p} > r_\mathrm{p,cr} = 0.16
 \frac{f_\mathrm{g}}{f_\mathrm{d}^2}
 \left(\frac{J_\mathrm{cr}}{0.1}\right)
 \left(\frac{\rho_\mathrm{p}}{3 \mathrm{g}/\mathrm{cm}^3}\right)^{-1}
 \left(\frac{t_\mathrm{disk}}{10^7\, \mathrm{years}}\right)^{-1}
\mathrm{mm}.
\end{equation}
When $f_\mathrm{g}=f_\mathrm{d}=3$, the critical radius in order for the secular GI to grow within $10^7\, \mathrm{years}$ is $r_\mathrm{p} \simeq 0.053\, \mathrm{mm}$. 

\section{Summary and Discussion \label{sec:conclusion}}
We investigated the secular GI for the case in which dust particles are very small.
A general analysis of the secular GI has been performed by \cite{Youdin2011}.
He used the hydrodynamic equation with the additional diffusion term.
His result may be applicable for even large dust particles.
However, his formulation is somewhat intuitive. We herein restrict ourselves to small dust particles and investigate the same instability rigorously.
We obtained essentially the same result, although the present study was conducted independently of \cite{Youdin2011}.
\cite{Youdin2011} considered the general turbulence, whereas we focused on the specific shear turbulence model reported by \cite{Sekiya1998} and discuss the condition for the secular GI in detail.

In the case of small particles, the stopping time is shorter than the turbulence auto-correlation time, $t_\mathrm{s} \ll t_\mathrm{e}$. We can neglect the particle inertia because the gas drag relaxation is very rapid. We derived an advection-diffusion equation for small particles in turbulence from the Langevin equation considering the external force field. We confirmed that the radial diffusion coefficient derived in the present paper is the same as that derived in \cite{Youdin2007}.

Considering the self-gravity of the particles with a thin disk approximation, we performed linear stability analysis of the advection-diffusion equation.
We calculated the growth rate of the secular GI.
Although the diffusion can stabilize the short wavelength mode, the long wavelength mode is always unstable, even though the turbulence is strong.
Next, we considered the effect of the finite thickness of the dust layer.
The thickness of the dust layer weakens the secular GI because of the lower density of a dust layer of finite thickness.
If the diffusion coefficient depends on the surface density, the growth rate changes.
In particular, if the diffusion coefficient decreases with the surface density, the growth rate of the secular GI can become very large.

Assuming the minimum-mass disk model, we calculated the growth rate of the secular GI.
We adopted the constant Richardson number density distribution model as the dust layer model \citep{Sekiya1998}. We calculated the turbulent diffusion coefficient from the density distribution assuming the mass flux balance and confirmed that the thin disk approximation is valid when the critical wavelength is smaller than the thickness of the dust layer.
If the dust-to-gas mass ratio is large or the temperature at the mid-plane is lower, the secular GI is faster than the radial drift due to the gas drag. In the case of the minimum-mass disk model, the radial drift is faster than the secular GI. The realistic surface density can be larger than that of the minimum-mass disk model. If the surface density is more than three times as large as that in the minimum-mass disk model, then the secular GI is faster than the radial drift.
In this case, the secular GI with millimeter-sized dust can develop within $5.3 \times 10^5$ years.

Note that the growth time is of the $e$-folding time scale, which is the time in which the perturbation quantities increase by a factor of $e$. Generally speaking, the planetesimal formation time should be larger than the growth time of the secular instability. 
The realistic planetesimal formation time depends on the amplitude of the initial perturbation and the nonlinear process that occurs after the linear growth.
Further investigation is necessary in order to clarify planetesimal formation by secular GI.

We restricted the present study to the strong drag case $t_\mathrm{s} \ll t_\mathrm{e}$ and confirmed that the analysis of \cite{Youdin2011} is valid when $t_\mathrm{s} \ll t_\mathrm{e}$.
His analysis may be applicable to more general parameters, such as $t_\mathrm{s} \gg t_\mathrm{e}$. Strictly speaking, we should solve the time evolution equation of the velocity distribution in this parameter regime, because, in general, the relaxation process of the velocity distribution is not negligible. In addition, viscosity may not be negligible. The hydrodynamic equation with a turbulent diffusion term used by \cite{Youdin2011} should be more rigorously justified. We will investigate the time evolution equation and the secular GI for more general cases in a future study.

In the linear stability analysis, we assumed the axisymmetric modes.
In general, non-axisymmetric modes can be excited.
In fact, non-axisymmetric density patterns were observed in previous studies \citep{Tanga2004,  Michikoshi2007, Wakita2008, Michikoshi2009, Michikoshi2010}.
These structures were interpreted as being similar to self-gravitating wakes induced by GI and Kepler shear \citep{Salo1995}. When we consider the gas-free model \citep{Michikoshi2007, Michikoshi2009} or when the gas drag force is weak \citep{Michikoshi2010}, the timescale of the GI is on the order of Kepler time. The timescale of the secular GI under strong gas drag and turbulent diffusion is much longer than Kepler time. Thus, the Kepler shear becomes significant relative to the GI. The shear makes a non-axisymmetric mode a tight trailing mode, which may eventually be smoothed by the turbulent diffusion.
Thus, the non-axisymmetric mode may not be able to grow sufficiently.
The axisymmetric mode is expected to grow in the secular timescale.

Axisymmetric high-density rings form due to the secular GI. As the secular GI develops, the dust-to-gas mass ratio increases, and the dust density at the mid-plane increases. 
Finally, the dust density at the mid-plane exceeds the Roche density when the dust-to-gas mass ratio becomes larger than $2\--20$ \citep{Sekiya1998, Youdin2002}.
When the dust surface density in the ring increases by a factor of $F$, the ring width decreases, as follows:
\begin{equation}
\Delta a \simeq \frac{\lambda_\mathrm{max}}{F} = 6.0 \times 10^{-3} f_\mathrm{d}^{-1} \left(\frac{F}{10}\right)^{-1} \left(\frac{J_\mathrm{cr}}{0.1}\right) \left(\frac{a}{\mathrm{AU}}\right)^{3/2} {\mathrm{AU}}, 
\end{equation}
where $\lambda_\mathrm{max} = 2 \pi / k_\mathrm{max}$.
On the other hand, the critical wavelength of the dynamical GI of the thin rotating disk is
\begin{equation}
\lambda_\mathrm{GI} = \frac{4 \pi^2 G \Sigma_\mathrm{d}}{\Omega^2} = 3.2 \times 10^{-4} f_\mathrm{d} \left(\frac{F}{10}\right) \left(\frac{a}{\mathrm{AU}} \right)^{3/2} \mathrm{AU}.
\end{equation}
The condition for $\Delta a \gg \lambda_\mathrm{GI}$ is written as 
\begin{equation}
f_\mathrm{d} F \ll 43.5 \left(\frac{J_\mathrm{cr}}{0.1} \right)^{1/2}.
\end{equation}
As shown in Figure \ref{fig:r_fgcrit}, the critical $f_\mathrm{g}$ value for $J_\mathrm{cr}=0.1$ at 1 AU is approximately 2.
Then, the critical wavelength of the dynamical GI is slightly smaller
than the shrunken width of the axisymmetric dense ring. We may be able to apply
the dynamical GI theory of the rotating disk.
In this case, the planetesimal mass and radius are estimated by the classical GI theory:
\begin{equation}
m_\mathrm{pl} \simeq \Sigma_\mathrm{d} \lambda_\mathrm{GI}^2 = 1.63 \times 10^{20} f_\mathrm{d}^3 \left(\frac{F}{10}\right)^3 \left(\frac{a}{\mathrm{AU}} \right)^{3/2} \mathrm{g},
\end{equation}
\begin{equation}
r_\mathrm{pl} \simeq  23.5 f_\mathrm{d} \left(\frac{F}{10}\right) \left(\frac{\rho_\mathrm{p}}{3 \mathrm{g}/\mathrm{cm}^{3}} \right)^{-1/3} \left(\frac{a}{\mathrm{AU}} \right)^{1/2} \mathrm{km}.
\end{equation}
The planetesimal size is larger than the classical planetesimal because the dust surface density is large.

In general, the factor $F$ may be larger than approximately $10$ and the initial value of $f_\mathrm{d}$ may be larger than $2$. Then, the classical GI theory of the thin rotating disk breaks down because of $\lambda_\mathrm{GI} > \Delta a$. In this case, stability analysis of a thin ring or filament of dust with turbulent stirring is necessary in order to estimate the planetesimal size. We intend to examine the stability of a thin filament of dust with turbulent stirring in a future study.

We thank Minoru Sekiya for helpful comments on this research. 
\bibliographystyle{apj}

\appendix

\section{Derivation of Coefficients \label{sec:derivation} }
We assume the timescale $T$ of the phenomenon consider herein to be much longer than the eddy turnover time $t_\mathrm{e}$, i.e., $T \gg t_\mathrm{e}$.
In the following discussion, the auto-correlation function is used in the integrals.
Since the auto-correlation is negligible for $t>t_\mathrm{e}$, the auto-correlation function described by Equation (\ref{eq:acf}) in the integrals is replaced by the delta function:
\begin{equation}
\phi(t) \simeq C \delta(t),
\end{equation}
where $\delta(t)$ is the Dirac delta function.
Assuming that the integral over the time $t$ is invariant,
\begin{equation}
\int C \delta(t) dt = \int v_\mathrm{t}^2 \exp \left(-\frac{|t|}{t_\mathrm{c}} \right) dt,
\end{equation}
we determine the coefficient of the delta function $C$ as follows:
\begin{equation}
C= 2 t_\mathrm{e} v_\mathrm{t}^2.
\end{equation}

The auto-correlation functions of the $x$ and $y$ components of the turbulence velocity are 
\begin{equation}
\phi_{xx}(t) \simeq 2 t_\mathrm{e} \sigma_{xx}^2 \delta(t) = 2 D_{\mathrm{g}xx} \delta(t),
\label{eq:cor1}
\end{equation}
\begin{equation}
\phi_{yy}(t) \simeq 2 t_\mathrm{e} \sigma_{yy}^2 \delta(t) = 2 D_{\mathrm{g}yy} \delta(t),
\label{eq:cor2}
\end{equation}
where $t_{\mathrm{t}x}$ and $t_{\mathrm{t}y}$ are the turbulence velocity dispersion,
$D_{\mathrm{g}xx}$ and $D_{\mathrm{g}yy}$ are the gas diffusion coefficients, where $D_{\mathrm{g}xx} = t_\mathrm{e} \sigma_{xx}^2 $ and $D_{\mathrm{g}yy} = t_\mathrm{e} \sigma_{yy}^2 $.
The cross-correlation function of the $x$ and $y$ components of the turbulent velocity is 
\begin{equation}
\phi_{xy}(t) \simeq 2 t_\mathrm{e} \sigma_{xy}^2 \delta(t)= 2 D_{\mathrm{g}xy} \delta(t),
\label{eq:cor3}
\end{equation}
where $\sigma_{xy}$ is the magnitude of the cross correlation, and $D_{\mathrm{g}xy} = t_\mathrm{e} \sigma_{xy}^2 $.

The formal solution of Equation (\ref{eq:s1}) is
\begin{equation}
\Delta x = \int_{t_0}^{t_0+\Delta t} \frac{dx}{dt} dt = \frac{\gamma^2 \int_{t_0} ^{t_0+\Delta t} v_{\mathrm{g}x} dt_1  + 2 \Omega \gamma \int _{t_0} ^{t_0+\Delta t}v_{\mathrm{g}y} dt_1 + \gamma \int _{t_0} ^{t_0+\Delta t} f_x dt_1 + 2 \Omega \int _{t_0} ^{t_0+\Delta t} f_y dt_1 }{\gamma^2 + \Omega^2}.
\end{equation}
The external forces $f_x$ and $f_y$ are assumed to be smooth functions with respect to $x$ and $y$, respectively.
The integrals of the external forces are approximated as follows:
\begin{equation}
\int _{t_0} ^{t_0+\Delta t} f_x dt_1 \simeq f_x \Delta t  
+ \frac{\partial f_x}{\partial x} \int_{t_0} ^{t_0+\Delta t} (x(t_1)-x(t_0)) dt_1,
\end{equation}
where we assume the change in position to be small.
If the gradient of the external force is small, the second term is negligible.
The integral of the external force is assumed to be as follows:
\begin{equation}
\int _{t_0} ^{t_0+\Delta t} f_x dt_1 \simeq f_x \Delta t.
\end{equation}

Therefore, the first-order solution is written as
\begin{equation}
\Delta x \simeq \frac{\gamma^2}{\gamma^2+\Omega^2} \int_{t_0} ^{t_0+\Delta t} v_{\mathrm{g}x} dt_1  +
 \frac{ 2\gamma \Omega}{\gamma^2+\Omega^2} \int_{t_0} ^{t_0+\Delta t} v_{\mathrm{g}y} dt_1  +
 \frac{\gamma f_x}{\gamma^2 + \Omega^2} \Delta t  +
 \frac{2 \Omega f_y}{\gamma^2 + \Omega^2} \Delta t. 
\label{eq:formal_sol1}
\end{equation}
In the same way, the formal solution of Equation (\ref{eq:s2}) is 
\begin{eqnarray}
\Delta y &\simeq& \int_{t_0}^{t_0+\Delta t} \frac{dy}{dt} dt = \int_{t_0}^{t_0+\Delta t} \left( v_y-\frac{3}{2} \Omega x \right) dt \nonumber \\
&=& - \frac{ \gamma \Omega}{2(\gamma^2+\Omega^2)} \int_{t_0} ^{t_0+\Delta t} v_{\mathrm{g}x} dt_1 + \frac{\gamma^2}{\gamma^2+\Omega^2} \int_{t_0} ^{t_0+\Delta t} v_{\mathrm{g}y} dt_1  + \frac{\gamma f_y}{\gamma^2 + \Omega^2} \Delta t  - \frac{\Omega f_x}{2(\gamma^2 + \Omega^2)} \Delta t \nonumber \\ 
&& - \frac{3}{2} \Omega x_0 \Delta t  - \frac{3}{2} \frac{\gamma^2\Omega }{\gamma^2+\Omega^2} \int_{t_0}^{t_0+\Delta t} dt_1 \int_{t_0}^{t_1} dt_2 v_{\mathrm{g}x} - 
\frac{ 3\gamma \Omega^2}{\gamma^2+\Omega^2} \int_{t_0}^{t_0+\Delta t}dt_1  \int_{t_0}^{t_1} dt_2 v_{\mathrm{g}y}.
\label{eq:formal_sol2}
\end{eqnarray}

Since the ensemble means of $v_{\mathrm{g}x}$ and $v_{\mathrm{g}y}$ are zero, the integrals are also zero:
\begin{equation}
\langle \int_{t_0} ^{t_0+\Delta t} v_{\mathrm{g}x} dt_1 \rangle = \int_{t_0} ^{t_0+\Delta t} \langle v_{\mathrm{g}x} \rangle dt_1 = 0,
\end{equation}
\begin{equation}
\langle \int_{t_0} ^{t_0+\Delta t} v_{\mathrm{g}y} dt_1 \rangle = \int_{t_0} ^{t_0+\Delta t} \langle v_{\mathrm{g}y} \rangle dt_1 = 0.
\end{equation}
Thus, the ensemble means of Equations (\ref{eq:formal_sol1}) and (\ref{eq:formal_sol2}) are 
\begin{equation}
\langle \Delta x \rangle \simeq  \frac{\gamma f_x}{\gamma^2 + \Omega^2} \Delta t  +
 \frac{2 \Omega f_y}{\gamma^2 + \Omega^2} \Delta t,
\end{equation}
\begin{eqnarray}
\langle \Delta y \rangle \simeq \frac{\gamma f_y}{\gamma^2 + \Omega^2} \Delta t  - \frac{\Omega f_x}{2(\gamma^2 + \Omega^2)} \Delta t - \frac{3}{2} \Omega x_0 \Delta t. 
\end{eqnarray}
We readily obtain the first-order coefficients as shown by Equations (\ref{eq;coef_01_x}) and (\ref{eq;coef_01_y}).

The ensemble mean of $\Delta x^2$ is
\begin{eqnarray}
\langle \Delta x^2 \rangle &\simeq& \frac{\gamma^4}{(\gamma^2+\Omega^2)^2} \int_{t_0} ^{t_0+\Delta t} dt_1 \int_{t_0} ^{t_0+\Delta t} dt_2 \langle v_{\mathrm{g}x}(t_1)  v_{\mathrm{g}x}(t_2) \rangle \nonumber \\
&&+ \frac{2\gamma^3 \Omega}{(\gamma^2+\Omega^2)^2} \int_{t_0} ^{t_0+\Delta t} dt_1 \int_{t_0} ^{t_0+\Delta t} dt_2 \langle v_{\mathrm{g}x}(t_1)  v_{\mathrm{g}y}(t_2) \rangle \nonumber \\
&&+ \frac{4\gamma^2 \Omega^2}{(\gamma^2+\Omega^2)^2} \int_{t_0} ^{t_0+\Delta t} dt_1 \int_{t_0} ^{t_0+\Delta t} dt_2 \langle v_{\mathrm{g}y}(t_1)  v_{\mathrm{g}y}(t_2) \rangle. 
\label{eq:xsq}
\end{eqnarray}
Since we adopt the white noise approximation, the auto-correlation functions are given by the Dirac delta functions as Equations (\ref{eq:cor1}), (\ref{eq:cor2}), and (\ref{eq:cor3}).
The integral in the first term of Equation (\ref{eq:xsq}) is evaluated as
\begin{equation}
\int_{t_0} ^{t_0+\Delta t} dt_1 \int_{t_0} ^{t_0+\Delta t} dt_2 \langle v_{\mathrm{g}x}(t_1)  v_{\mathrm{g}x}(t_2) \rangle = \int_{t_0} ^{t_0+\Delta t} dt_1 \int_{t_0} ^{t_0+\Delta t} dt_2 2  D_{\mathrm{g}x} \delta(t_1 - t_2) =  2 D_{\mathrm{g}x} \Delta t.
\end{equation}
Thus, we obtain the diffusion coefficients $D_{xx}$ as Equation (\ref{eq:difx}).

The ensemble mean of $\Delta y^2$ is
\begin{eqnarray}
\langle \Delta y^2 \rangle &\simeq& \frac{\gamma^2 \Omega^2}{4(\gamma^2+\Omega^2)^2} \int_{t_0} ^{t_0+\Delta t} dt_1 \int_{t_0} ^{t_0+\Delta t} dt_2 \langle v_{\mathrm{g}x}(t_1)  v_{\mathrm{g}x}(t_2) \rangle \nonumber \\
&& - \frac{\gamma^3 \Omega}{(\gamma^2+\Omega^2)^2} \int_{t_0} ^{t_0+\Delta t} dt_1 \int_{t_0} ^{t_0+\Delta t} dt_2 \langle v_{\mathrm{g}x}(t_1)  v_{\mathrm{g}y}(t_2) \rangle \nonumber \\
&& + \frac{\gamma^2 \Omega^2}{(\gamma^2+\Omega^2)^2} \int_{t_0} ^{t_0+\Delta t} dt_1 \int_{t_0} ^{t_0+\Delta t} dt_2 \langle v_{\mathrm{g}y}(t_1)  v_{\mathrm{g}y}(t_2) \rangle \nonumber \\
&& + A. 
\end{eqnarray}
where
\begin{eqnarray}
A &=& \biggl\langle 2 \left( - \frac{3}{2} \frac{\gamma^2\Omega }{\gamma^2+\Omega^2} \int_{t_0}^{t_0+\Delta t} dt_1 \int_{t_0}^{t_1} dt_2 v_{\mathrm{g}x} - \frac{ 3\gamma \Omega^2}{\gamma^2+\Omega^2} \int_{t_0}^{t_0+\Delta t} dt_1 \int_{t_0}^{t_1} dt_2 v_{\mathrm{g}y} \right) \nonumber \\
&& \times \left( - \frac{ \gamma \Omega}{2(\gamma^2+\Omega^2)} \int_{t_0} ^{t_0+\Delta t} v_{\mathrm{g}x} dt_1 + \frac{\gamma^2}{\gamma^2+\Omega^2} \int_{t_0} ^{t_0+\Delta t} v_{\mathrm{g}y} dt_1  \right) \nonumber \\
&& + \left(  - \frac{3}{2} \frac{\gamma^2\Omega }{\gamma^2+\Omega^2} \int_{t_0}^{t_0+\Delta t} dt_1 \int_{t_0}^{t_1} dt_2 v_{\mathrm{g}x} - \frac{ 3\gamma \Omega^2}{\gamma^2+\Omega^2} \int_{t_0}^{t_0+\Delta t} dt_1 \int_{t_0}^{t_1} dt_2 v_{\mathrm{g}y} \right)^2 \biggr\rangle.
\end{eqnarray}
The expansion of $A$ contains multiple integrals, as follows:
\begin{equation}
\int _{t_0} ^{t_0+ \Delta t} dt_3 \int _{t_0} ^{t_0+ \Delta t} dt_1 \int _{t_0} ^{t_1} dt_2 \delta(t_2- t_3) = \frac{1}{2} \Delta t^2,
\end{equation}
\begin{equation}
\int _{t_0} ^{t_0+ \Delta t} dt_1 \int_{t_0} ^{t_1} dt_2 \int _{t_0} ^{t_0+ \Delta t} dt_3 \int _{t_0} ^{t_3} dt_4 \delta(t_2- t_4) = \frac{2}{3} \Delta t^3.
\end{equation}
These integrals are higher-order terms.
Accordingly, the term A is $\sim O(\Delta t^2)$, which is negligible. 
The diffusion coefficients $D_{yy}$ are then obtained as shown in Equation (\ref{eq:dify}).
\end{document}